ADAPTIVE TIME-DELAY HYPERCHAOS SYNCHRONIZATION IN LASER DIODES SUBJECT TO OPTICAL FEEDBACK

E. M. Shahverdiev [1] and K. A. Shore [2]

School of Informatics, University of Wales, Bangor, Dean Street, Bangor, LL57 1UT, Wales, UK

ABSTRACT

In this paper a proposal is made of an adaptive coupling function for achieving synchronization between two lasers subject to optical feedback.Such a control scheme requires knowledge of the systems' parameters. We demonstate that when these parameters are not available on-line parameter estimation can be applied.

1.INTRODUCTION

The seminal papers by Pecora and Carroll [1 ] and Ott, C.Grebogi and J.A.Yorke [2] on chaos synchronization have stimulated a wide range of research activity : a recent comprehensive review of such work is found in the focussed issues on chaos control [3] and references therein. Application of chaos control theory can be found in secure communications,optimization of nonlinear system performance and modeling brain activity and pattern recognition phenomena [3]. A particular focus of the work being the development of secure optical communications systems based on control and synchronization of laser chaos [4-7].For practical use of this approach particular emphasis is given to the use of chaotic semiconductor lasers see eg [7]. It has been shown [8] that security cannot be guaranteed in a communications format using simple chaotic systems - ie those with a single positive Lyapunov exponent. It is thus appreciated that to obtain reliable communications systems attention should be directed at hyperchaotic systems - ie those with with two or more positive Lyapunov exponents .It has been claimed previously that the number of driving variables needed for synchronization in case of hyperchaotic systems should be equal to the number of positive Lyapunov exponents [3]. However, such a requirement is highly undesirable in communication applications, as most communication schemes use just one signal for transmission [3]. More recently it was argued [9-10 ], that hyperchaos control is possible using fewer driving variables than the number of positive Lyapunov exponents [9], and indeed even with zero- driving variables using the method of parameter change advocated in [10]. Moreover , it has been shown recently that hyperchaos control is possible with a single variable even in the case of time delay systems, when the number of positive Lyapunov exponents ,in principle, can be infinite [11].This result is of particular importance for the use of external cavity laser diodes for chaotic optical communications [ 7]. In addition to applications in communications the implications of the study of synchronization phenomenon in time-delayed systems can be considered as a special case of spatio-temporal chaos control. Time-delay systems are infinite-dimensional and more interestingly by changing the time-delay one can obtain different numbers of the positive Lyapunov exponents [11 ].

---

[1]Permanent address: Institute of Physics, 370143 Baku,Azerbaijan
[2]Electronic address: alan@sees.bangor.ac.uk

In [11] and [12] use is made of both uni-directional and bi-directional couplings between the master and slave time-delay systems.Then an estimate is made of the coupling strength, for the given coupling function, needed for the synchronization between the drive and response time-delay systems.Usually two dynamical systems are termed synchronized if the differerence between their states converges to zero for $t \to \infty$ [1-2].Recently [13-14], a generalization of this concept was proposed, where two systems are termed as being synchronized if a functional relation exists between the states of both systems.

In this paper we propose a general adaptive coupling (linking) function needed for synchronization between two time-delay systems.Thus we are not imposing threshold restrictions on the coupling strength.

Laser systems with optical feedback are prominent representatives of time-delay systems[6-7, 15-16].In the light of this we also apply the proposed approach to the case of synchronization between two lasers subject to optical feedback. Our results show that one can use a time delay coupling function to accomplish synchronization between the laser systems and this synchronization method is different from that of [15, 17]. We argue that such a diversity allows for more flexibility in practical control problems.

Usually, in the context of nonlinear dynamical systems, the method of adaptive control applies a feedback loop in order to drive the system parameter (or parameters) to the values required so as to achieve a target state. This is achieved by adding the equation for the evolution of the parameter(s) to the evolution dynamics of the dynamical systems [18-29]. Such a scheme is adaptive, because the parameters which determine the nature of the dynamics self-adjust or adapt themselves to yield the desired dynamics.

In this paper we use the term 'adaptive' in a slightly different sense. The proposed method of chaos synchronization between two chaotic systems can also be interpreted as follows: we apply a control law to the process model to reach the reference model (desired or target state), which in principle can be entirely different from the process model not only due to parameter(s) mismatches, but also by structure and/or dynamics; in other words, the task is to design a control force and apply it to the process model to reach an entirely different targeted state. Such a scheme is also adaptive, as in the above procedure the linkage function depending on the nature of the systems' dynamics, structure self-adjuct or adapt themselves to yield the desired dynamics [21]. By definition, the adaptive principle is remarkably robust and and efficient in generic nonlinear systems and may therefore be of immense utility in a large variety of practical control problems.

## 2. ADAPTIVE COUPLING FOR CHAOS SYNCHRONIZATION

Following [11-12] we write the time-delay system under consideration in the form:

$$\frac{dx}{dt} = f(x, x_\tau),$$

$$\frac{dy}{dt} = g(y, y_\tau) + W(x, y), \quad (1)$$

where $f$ and $g$ are arbitrary time delay functions such that the corresponding dynamics exhibit chaotic behavior; $x_\tau := x(t - \tau)$, $y_\tau := y(t - \tau)$; where $\tau$ is the time-delay; the term $W(x, y)$ in equation (1) is responsible for coupling (linkage) between the master (driving) (first equation in (1)) and slave (response) (second equation in (1)) systems. The relaxation terms proportional to $x$ and $y$, usually written separately in the right hand sides of the system (1), are incorporated into the f and g. In addition, one have to keep in mind that in general the dynamical variables and $f, g$ and $W$ can be high dimensional vectors.

Let us choose the adaptive coupling function (or control input) $W(x, y)$ in the system (1) as:

$$W(x, y) = f(x, x_\tau) - g(y, y_\tau) - Q(x, y), \quad (2)$$

with $Q(x, y)$ such that error dynamics of $x - y = e$ will be stable. For $Q(x, y) = B(x - y) = Be$ with negative $B$ we obtain:

$$\frac{de}{dt} = Be, \quad (3)$$

whose solution decays exponentially. Generalization to the case of different delay functions with different time delays $\tau_1$ and $\tau_2$ is straightforward. In this case one can choose the coupling function

$$W(x, y) = f(x, x_{\tau_1}) - g(y, y_{\tau_2}) - Q(x, y). \quad (4)$$

The idea behind the way of choosing the adaptive coupling (or control input) to achieve synchronization is to cancel the nonlinear terms of the system. We want to linearize the system to make it more tractable and to use linear control theory. Sometimes this approach is called feedback linearization, see for more details and pitfalls, e.g. [22] and references therein.

As indicated above many laser communication systems are prominent representatives of time-delay systems. In this work we apply the proposed approach to the case of synchronization between two semiconductor lasers subject to optical feedback. There can be different types of couplings between the slave and master systems.For example in [15] the light that is injected into the slave system is included in the equations in a way similar to the light coming from the external resonator.This approach is widely used to describe the effects of coherent light injection into semiconductor lasers.In this paper we propose a new type of coupling between master and slave systems to achieve synchronization between these systems.A general form for synchronization condition is obtained from a consideration of the following systems of the Lang-Kobayashi equations [15,17] for the real electric field amplitude E(t), slowly varying phase $\Phi(t)$ and the carrier number n(t) for the:master (with subscript M),

$$\frac{dE_M}{dt} = \frac{1}{2}Gn_M E_M + k_M E_M(t-\tau)\cos(\omega_0\tau + \Phi_M(t) - \Phi_M(t-\tau)),$$

$$\frac{d\Phi_M}{dt} = \frac{1}{2}\alpha Gn_M - k_M \frac{E_M(t-\tau)}{E_M(t)}\sin(\omega_0\tau + \Phi_M(t) - \Phi_M(t-\tau)),$$

$$\frac{dn_M}{dt} = (p-1)J_{th} - \gamma n_M(t) - (\Gamma + Gn_M)E_M^2, \tag{5}$$

and slave lasers (with subscript S),

$$\frac{dE_S}{dt} = \frac{1}{2}Gn_S E_S + k_S E_S(t-\tau)\cos(\omega_0\tau + \Phi_S(t) - \Phi_S(t-\tau)) + W,$$

$$\frac{d\Phi_S}{dt} = \frac{1}{2}\alpha Gn_S - k_S \frac{E_S(t-\tau)}{E_S(t)}\sin(\omega_0\tau + \Phi_S(t) - \Phi_S(t-\tau)),$$

$$\frac{dn_S}{dt} = (p-1)J_{th} - \gamma n_S(t) - (\Gamma + Gn_S)E_S^2, \tag{6}$$

coupled by the linkage function

$$W = K_W(E_M - E_S) + \frac{1}{2}G(n_M E_M - n_S E_S) + k_M E_M(t-\tau)\cos(\omega_0\tau + \Phi_M(t) - \Phi_M(t-\tau))$$

$$- k_S E_S(t-\tau)\cos(\omega_0\tau + \Phi_S(t) - \Phi_S(t-\tau)), \tag{7}$$

where $G$ is the diffential optical gain;$\tau$ is the master laser's external cavity round-trip time;$\alpha$- the linewidth enhancement factor;$\gamma$- the carrier density rate;$\Gamma$-the cavity decay rate;$p$-the pump current relative to the threshold value $J_{th}$ of the solitary laser;$\omega_0$ is the angular frequency of the solitary laser;$k$ is the feedback rate;$K_W$ is the coefficent determining the speed of achieving synchronization between the master and slave lasers.

One can see easily that with for type of coupling with positive $K_W$ that the difference signal $e_E = E_M - E_S$ approaches zero, as the error dynamics in this case obey the following equation:

$$\frac{de_E}{dt} = -K_W e_E. \tag{8}$$

(Throughout this paper we introduce the relaxation or damping term to overcome the necessity of identical inital conditions in the coupled master and slave laser systems.)

In the above scheme of synchronization the master and slave systems' parameter, namely the gain was the same for both systems. Generalization of the coupling function to the case of laser systems with different parameters is straightforward; for example, with different gain parameters the coupling function is:

$$W = K_W(E_M - E_S) - \frac{1}{2}(G_M n_M E_M - G_S n_S E_S) - k_M E_M(t-\tau)\cos(\omega_0 \tau + \Phi_M(t) - \Phi_M(t-\tau))$$

$$+ k_S E_S(t-\tau)\cos(\omega_0 \tau + \Phi_S(t) - \Phi_S(t-\tau)). \tag{9}$$

As was pointed out in [23], in many representative cases, chaos synchronization can be understood from the existence of a global Lyapunov function of the difference signals.In other words, the global asymptotic stability can be investigated by the Lyapunov function approach [22]. For error dynamics $e_E$ (8), one can use the Lyapunov function

$$L = e_E^2. \tag{10}$$

As

$$\frac{dL}{dt} = -K_W e_E^2, \tag{11}$$

can be made strictly negative for positive $K_W$ (except for $e_E = 0$) we conclude that the asymptotic stability is global.

Thus based on our recent results we have several possibilities for achieving synchronization between chaotic laser diodes:according to [17] if the coupling between master and slave systems is of the form

$$W = \sigma E_M(t-\tau_c)\cos(\omega_0 \tau_c + \Phi_S(t) - \Phi_M(t-\tau_c)), \tag{12}$$

(where $\sigma$ is the coupling strength between the master and slave lasers; $\tau_c$ is the light propagation time from the right facet of the master laser to the right facet of the slave laser) then the synchronization condition is

$$k_M = k_S + \sigma. \tag{13}$$

Here in this paper we have proposed another type of linkage function (7) for the synchronization purposes without strict condition on the systems' parameters.

3. ADAPTIVE SYNCHRONIZATION WITH UNKNOWN PARAMETERS

The systems parameters from eqs.(5-6) are required for the adaptive synchronization coupling function. In the case that these parameters are not available, one can apply the on-line parameter estimation method. In principle the number of unavailable parameters can be equal to the total number of systems' parameters. However in this paper we confine ourselves to the case of single parameter estimation, namely to gain estimation. So let us suppose that the gain's estimated value $G_1$ is different from the required (for synchronization) gain value $G$. With the estimated value of gain the adaptive coupling function would be of the form

$$W = K_W(E_M - E_S) + \frac{1}{2}G_1(n_M E_M - n_S E_S) + k_M E_M(t - \tau)\cos(\omega_0\tau + \Phi_M(t) - \Phi_M(t - \tau))$$

$$-k_S E_S(t - \tau)\cos(\omega_0\tau + \Phi_S(t) - \Phi_S(t - \tau)), \tag{14}$$

Under these conditions it is easy to verify that error dynamics now will satisfy the following equation:

$$\frac{de}{dt} = -K_W e - \frac{1}{2}(n_M E_M - n_S E_S)(G - G_1), \tag{15}$$

In other words, for $G \neq G_1$ the error $e$ will not approach zero, as required for synchronization purposes. The situation can be rectified, if we add the following equation for the parameter estimation error $e_G = G - G_1$ to the previous equation (15):

$$e_G = G_1 - G = e\frac{1}{2}(n_M E_M - n_S E_S) \tag{16}$$

Now we shall demonstrate the the origin of the systems (15)-(16) is asymptotically stable,i.e. that the synchronized state is asymptotically stable. Indeed, by choosing the following Lyapunov function:

$$L = \frac{1}{2}(e^2 + e_G^2), \tag{17}$$

it is trivial to check that

$$\frac{dL}{dt} = -K_W e^2 + \frac{1}{2}(n_M E_M - n_S E_S)e_G e - \frac{1}{2}(n_M E_M - n_S E_S)e_G e = -K_W e^2 < 0. \qquad (18)$$

Thus we demonstrate that the stability of the adaptive control law with unknown parameters is asymptotic.

The practical implementation of the proposed scheme can be based on the approaches developed in [24-26]. In these papers it has been shown that a scheme of chaos control for external cavity laser diodes can be effected where a periodic state of the dynamics is selected from the chaotic dynamics.The key to that approach is the utilisation of an error signal which defines the difference between the chaotic state and the targeted state. As the target state is approached the generated error signal reduces to zero. It was shown in those papers that optoelectronic feedback provides a straightforward means for generating the requiste error signal [24-26]. It is noted that the linkage function defined in the present work can be expressed as an error signal between the dynamics of the master and slave external cavity lasers. As the linkage function brings the dynamics of the two laser systems into synchronization the corresponding error signal will diiminish to zero. One approach to the practical implementation of the synchronization of the present scheme would thus again be based on the use of the optoelectronic feedback.

In other words, for the practical realization of the synchronization scheme we essentially inject the amplified difference signal between the master and slave lasers' outputs to the slave laser. Comparing our approach with other widely known methods we notice that in general the present synchronization procedure is different from that of [15,17] and we argue that it offers more flexibility in practical control problems.

4.CONCLUSION

In this paper we show how one can synchronize two chaotic time delay systems in the general case by choosing an appropriate delay adaptive coupling function.We apply the proposed approach to the case of synchronization between two semiconductor lasers subject to optical feedback. We also demonstrate that when the parameters of the systems to be synchronized are not available, then on-line parameter estimation can be applied.

5.ACKNOWLEDGEMENTS

This work is supported by the UK Engineering and Physical Sciences Research Council grant GR/R22568/01.